\documentclass[12pt,tightenlines,eqsecnum,floats,aps,amsmath,amssymb,nofootinbib,prd,shownopacs,floatfix]{revtex4}

\usepackage{setspace}
\usepackage{amssymb}
\usepackage{graphicx,wrapfig}
\usepackage{enumerate} 
\usepackage{color}
\usepackage{mathrsfs}
\usepackage{subfigure}
\definecolor{mg}{rgb}{0.0, 0.5, 0.0}

\def\be{\nopagebreak[3]\begin{equation}}
\def\ee{\end{equation}}
\def\ba{\nopagebreak[3]\begin{eqnarray}}
\def\ea{\end{eqnarray}}
\newcommand{\f}{\frac}
\def\rmd{{\rm d}}
\def\lp{\ell_{\rm Pl}}
\def\g{\mathfrak{g}}
\def\P{\mathcal{P}}

\def\t{\tilde}

\def\h{\hat}

\def\db{\delta_b}
\def\dc{\delta_c}
\def\T{\mathcal{T}}

\begin{document}

\title{Properties of a recent quantum extension of the Kruskal geometry}

\author{Abhay Ashtekar$^{1}$, Javier Olmedo$^{2,3}$, }
\affiliation {
1. Institute for Gravitation and the Cosmos, Penn State University,
University Park, PA 16801 \\
2. Department of Physics and Astronomy, Louisiana State University,
Baton Rouge, LA 70803\\
3. Departamento de F\'isica Te\'orica y del Cosmos, Universidad de Granada,  Granada-18071, Spain }
\begin{abstract}

Recently it was shown that, in an effective description motivated by loop quantum gravity, singularities of the Kruskal space-time are naturally resolved \cite{aoslett,aos}. In this note we explore a few  properties of this quantum corrected effective metric. In particular,  we (i)  calculate the Hawking temperature associated with the horizon of the effective geometry and show that the quantum correction to the temperature is completely negligible for macroscopic black holes, just as one would hope;  (ii) discuss the subtleties associated with the asymptotic properties of the space-time metric, and show that the metric is asymptotically flat in a precise sense; (iii)  analyze the asymptotic fall-off of curvature; and, (iv) show that the ADM energy is well-defined (and agrees with that determined by the horizon area), even though the curvature falls off less rapidly than in the standard asymptotically flat context.
\end{abstract}
\maketitle

\section{Introduction}
\label{s1}

It has long been suggested, especially by John Wheeler, that the reason why singularities can result from dynamics of general relativity is simply that the theory ignores quantum effects. However, in recent years, there has been some debate as to whether inclusion of quantum gravity effects by itself would be sufficient. For example,  motivated by the AdS/CFT correspondence,  it has been argued that quantum gravity will/should \emph{not} resolve certain (bulk) singularities, including those of  the classical  Schwarzschild-Anti-de Sitter space-times \cite{eh}.  More recently,  it was found that plausible counter examples to cosmic censorship \cite{weakgravity1,weakgravity2} would be removed if the weak gravity conjecture (WGC) holds. These results are then sometimes interpreted as implying that a quantum gravity theory that does not include some version of the WGC would not be physically viable. Wheeler's argument, on the other hand, suggests otherwise: violation of cosmic censorship in classical general relativity may simply be an artifact of ignoring the quantum nature of geometry.  In this view, the focus shifts away from possible constraints on matter to make the \emph{classical} theory satisfactory, to whether the \emph{quantum} theory is singularity-free and predictive. 

Loop quantum gravity (LQG) provides a concrete illustration of how the quantum nature of geometry could play the crucial role. The theory has a fundamental discreteness, encapsulated in an area gap $\Delta$ --the lowest non-zero eigenvalue of the area operator-- that dictates the quantum corrections to Einstein's equations. These modifications have been worked out in great detail in loop quantum cosmology (LQC)  (see, e.g., \cite{asrev,ps,iaps}). In this context, Einstein's equations provide an excellent approximation to LQC away from the Planck regime but  then the  LQC corrections dominate once curvature approaches the Planck scale, resolving the singularity. In particular, these modifications naturally provide upper bounds to curvature scalars that go as  inverse powers of $\Delta$. Therefore, while all physical quantities are bounded in the quantum theory, since  $\Delta \to 0$ in the classical limit,   the upper bounds go to infinity and we are back to classical singularities.  In all cases studied in detail so far, it is the quantum nature of geometry that plays the decisive role in making the quantum theory singularity-free, without any additional  constraints on matter fields. 

It is natural to ask whether the black hole singularities are also naturally resolved by the same quantum gravity effects. The simplest context is provided by the Schwarzschild-Kruskal space-time. To analyze whether this singularity is resolved, it suffices to restrict oneself to the black hole region that is bounded by  the singularity in the future and event horizons in the past, often referred to as the \emph{Schwarzschild interior}.  Now,  since this interior is isometric to the (vacuum) Kantowski-Sachs cosmological model, one can adopt procedures that have been used to analyze  homogeneous but anisotropic cosmologies in LQC.  Therefore, there has been considerable work on the Schwarzschild interior over the last 15 years or so  (see, e.g.   \cite{ab,lm,bv,dc,ck,cgp,bkd,djs,cs,oss,cctr,yks}  for investigations relevant for this note).  In all these treatments, the space-like singularity was resolved but detailed examination showed that the resulting quantum extension of the classical Schwarzschild  interior had some undesirable or puzzling features. For example,  one generally begins by introducing certain fiducial structures in the construction of the classical phase space for mathematical convenience, with the expectation that the final physical results would be independent of the specific choices made.  This was not always the case (see, e.g., \cite{ab,lm,cgp}).  While the final results in \cite{bv,dc,cs,oss} are free of this drawback, one finds that there are large quantum gravity effects in low curvature regions. For example, for large black holes, the quintessentially quantum transition from a trapped to an anti-trapped region can occur in  regions with arbitrarily small curvature in \cite{cs}, while quantum dynamics drives the effective trajectories to regions of phase space where the basic underlying assumptions are violated in \cite{bv,dc,cctr}.  (A succinct discussion of all these limitations can be found in Section IV.D of \cite{aos}.) This situation was analyzed in some detail more recently and it was shown that all known limitations of the LQG description of Schwarzschild interior can be overcome by suitably modifying a key step in the quantization procedure \cite{aoslett,aos}.  

This recent analysis also extended the previous work in two directions. First, the emphasis in the previous works was generally on issues rooted in anisotropic cosmologies, suggested by the Kantowski-Sachs space-time --such as bounces of various scale factors \cite{bv,bkd,cs,djs,cctr,dc2}, behavior of the energy density, expansion scalar,  shear potentials of the Weyl curvature {\cite{js}}, and, geodesic completeness and generic resolution of strong singularities \cite{ss}.  By contrast, the analysis in \cite{aoslett,aos} focused on issues that are central to black holes, such as trapped and anti-trapped regions, black hole type and white hole type horizons,  and the behavior of the static Killing field as one passes from the original Kruskal space-time to its quantum extension.  Second, while the previous analysis was confined to the Schwarzschild interior, Refs \cite{aoslett,aos} also contained a specific proposal to extend the quantum corrected, effective description to the asymptotic regions (see also \cite{hlkn}). 

The purpose of this note is to discuss a few properties of the effective space-time constructed in \cite{aos} that shed further light on its viability.  In broad terms, there are three aspects of the effective geometry of interest:  (i) singularity resolution; (ii) near horizon geometry; and (iii) asymptotic behavior.  While all three aspects were discussed in Refs \cite{aoslett,aos},  the focus was on the possibility of overcoming limitations of previous investigations which were confined to the Schwarzschild interior. Therefore,  issues related to the singularity resolution were analyzed in detail. In particular, it was shown that not only do curvature invariants remain bounded in the quantum corrected effective space-time, but the bounds are \emph{universal}. More precisely, to the leading order,  upper bounds on the curvature scalars $R^{2}$,\, $R_{ab} R^{ab}$,\, $R_{abcd} R^{abcd}$ are numerical multiples of $1/\Delta^{2}$, where $\Delta$ is the area gap, showing that the origin of the singularity resolution lies firmly in the quantum nature of geometry. The bounds are universal in the sense that they are insensitive to the mass of the macroscopic black holes considered. This upper bound is reached on the \emph{transition surface}  ${\mathcal{T}}$ which separates the trapped (i.e., black hole type) region, which the effective metric shares with the classical metric, and the untrapped (i.e. white hole type) region that represents the quantum extension.  This surface $\mathcal{T}$ replaces the classical singularity. In the classical limit, we have $\Delta \to 0$ whence the upper bound diverges, and we have a singularity in place of the transition surface. These general features are the same as in the resolution of the big-bang singularity in LQC,  where the bounce-surface replaces $\mathcal{T}$, and the upper bound on curvature is again governed  only by the area gap, being insensitive to the specific matter content of the universe. Thus, there is a certain underlying unity in the manner in which space-like singularities are resolved due to quantum geometry effects.

In this note we further investigate aspects  (ii) and (iii).  Specifically, although the  analytical form of the effective metric is quite complicated, one can still calculate the temperature associated with the black hole using Euclidean methods (see, e.g. \cite{fulling}).  
Thus, one exploits the fact that the temperature associated with  a thermal state of a (test) quantum field is succinctly captured by the periodicity in imaginary time of the Green's function of that field \cite{ggmp}. We will find that the quantum correction to the Hawking temperature is minute for macroscopic black holes; {it is $\mathcal{O}(10^{-106})$ for a solar mass black hole!} In \cite{aos} it was found that  the quantum modification of the near horizon curvature  is extremely small for macroscopic black holes. The result on the Hawking temperature translates those considerations to more direct physical terms.

Aspect (iii) --the  asymptotic behavior of the effective geometry-- was not analyzed in as much detail as the first two aspects. This was largely because --as discussed in sections \ref{s3} and \ref{s4}--  the effective metric is in excellent agreement with the classical theory near horizons, and calculations reported in \cite{aos} showed that the agreement improves as one moves away from the horizons in the outward direction. Therefore, it was assumed that the agreement with the Schwarzschild geometry would continue to get better as one recedes even further.  However, soon afterwards it was realized that the issue of comparison requires greater care.
The effective metric $g_{ab}$ \emph{does} approach a flat metric (``as $1/r$'') --the precise sense is spelled out in Section \ref{s4}--  but the curvature invariants do not fall-off as fast as they do in standard  treatments of asymptotics (e.g., \cite{geroch-cinn, ah,aa-ein,wald}). However, it is well-known that the ADM energy can remain well-defined even if the asymptotic fall-off is much slower than what is generally assumed  (see, e.g., \cite{schoen}). We will show that the same is true for the weaker asymptotic fall-off of the effective metric.  More precisely, the situation is as follows. With the standard fall-off, various expressions of the ADM energy --those involving just the spatial metric, those involving the Ricci curvature of this metric, the one in terms of the electric part of the Weyl curvature, the one involving asymptotic time-translational Killing field, etc -- all agree \cite{aaam1,aaam2}. This is no longer true if we have a weaker fall-off and the ADM energy has to be defined with greater care.  When this is done using the effective metric, the ADM energy is well-defined and its value is the expected one. Therefore, although the asymptotic curvature of the effective metric  falls-off more slowly than that of the Schwarzschild metric of classical general relativity, the main conclusions of \cite{aoslett,aos} still hold.  Furthermore, as we show,  to find a discernible deviation from the Schwarzschild metric for macroscopic black holes (for which the effective theory  was developed) one would have to move away from the black hole a distance that is orders of magnitude greater than the radius of the observable universe.

The paper is organized as follows. In Section \ref{s2}, we recall the effective metric and discuss some of its properties. In Section \ref{s3}, we discuss the Hawking temperature and  in Section \ref{s4}, asymptotic properties of the quantum corrected effective metric. In Section \ref{s5} we summarize the main results as well as work by others that extends the findings of \cite{aoslett,aos}, and point out limitations of this effective description  which in turn suggest directions in which it can be improved. We also have taken this opportunity to address two comments \cite{brahmacomment,mb-comment} on the results of \cite{aos}. Specifically,  in  Sections \ref{s4.2} and \ref{s5} we explain why the concerns expressed there are misplaced.

\section{Effective Equations and the near horizon geometry}
\label{s2}

For convenience of the reader, we will begin by recalling the effective metric obtained in \cite{aoslett,aos}, first  in the interior region bounded by the trapping (i.e. black hole type) and anti-trapping (i.e. white hole type) horizons, and then in the exterior region. We then introduce Eddington-Finkelstein coordinates in neighborhoods of these horizons to make it explicit that the interior metric is smoothly related to the exterior.

\subsection{Effective metric}
\label{s2.1}

As is common in LQC, the effective solution was first obtained in a symmetry reduced phase space framework and then expressed as a quantum corrected space-time metric.  The interior region is foliated by a family of space-like \emph{homogeneous} 3-surfaces and the exterior region, by a family of \emph{homogeneous} time-like surfaces (in both cases the metric on the 3-surfaces is anisotropic). Because of homogeneity,  the phase space $\Gamma$ is only 4 dimensional, coordinatized by $b, c;\, p_{b}, p_{c}$ in the interior region and their tilde version in the exterior. Quantum corrections are encapsulated in two `quantum parameters' $\delta_{b}, \delta_{c}$ (that are the same in both regions, but the subscripts carry a tilde in the exterior region just for notational consistency). Their specific form is motivated by the fact that in LQG the curvature operator is constructed by dividing the holonomy around suitable closed plaquettes by the area enclosed by those plaquettes, and then shrinking the loop till it attains the minimum non-zero eigenvalue of the area operator, $\Delta \approx 5.17 \lp^{2}$,  in LQG.  In the present approach \cite{aoslett,aos}, the plaquettes are chosen to lie on the transition surface $\mathcal{T}$ that replaces the classical singularity, and separates the trapped region from the anti-trapped region in the quantum extension of the Schwarzschild interior. The quantum parameters are then given by:
\be\label{db-dc}
\db=\Big(\frac{\sqrt{\Delta}}{\sqrt{2\pi}\gamma^2m}\Big)^{1/3} \qquad {\rm and} \qquad
L_{o}\dc=\frac{1}{2} \Big(\frac{\gamma\Delta^2}{4\pi^2 m}\Big)^{1/3},
\ee
where $\gamma \approx 0.2375$ is the Barbero-Immirzi parameter of LQG and $L_{o}$ is an infrared regulator introduced to make the phase space description well-defined.%
\footnote{None of the final physical results depend on $L_{o}$. They depend only on combinations of $b, \,L_{o}^{-1}c, \, L_{o}^{-1}p_{b},\, p_{c}$, which are all independent of the choice of  $L_{o}$.}
In the classical limit both quantum parameters go to zero and one recovers the standard Schwarzschild geometry; it is the quantum geometry induced corrections involving $\delta_{b}, \, \delta_{c}$ that resolve  the singularities of the Kruskal space-time. Finally, the parameter $m$ that characterizes the solution is a constant of motion, i.e., a Dirac observable, of the effective solution.
In the interior region, the effective metric is given by:
\be\label{intmetric}
g_{ab} \rmd x^{a} \rmd x^{b} \equiv \rmd s^2 = - \f{\gamma^{2} p_{c} \delta_{b}^{2}}{\sin^{2} (\delta_{b} b)}\, \rmd T^2 + \f{p_b^2}{p_c L_o^2}\, \rmd x^2 + p_c \,\rmd \omega^{2} \, ,\ee
where $\rmd \omega^{2}$ is the metric on a unit 2-sphere.  $T$ plays the role of time, and the translational Killing vector is $\partial/\partial x$; thus the phase space variables depend only on $T$. One can think of $x$ as the standard  Schwarzschild coordinate $t$ which is space-like in the interior region, and $(2m)\, e^{T}$ as the Schwarzschild coordinate $r$. 

Solutions of the effective equations are given by:
\begin{align}
&\tan\Big(\frac{\delta_c{c}\left(T\right)}{2}\Big)=\frac{\gamma L_0\delta_c}{8m}e^{-2T}\, ,\qquad
{p}_c\left(T\right)=4m^2\Big(e^{2T}+\frac{\gamma^2L_0^2\delta_c^2}{64m^2}e^{-2T}\Big)\,,\label{cpc}\\
&\cos\left(\delta_b{b}\left(T\right)\right)=b_0\tanh\Big(\frac{1}{2}\big(b_0T+2\tanh^{-1} (b_0^{-1})\big)\Big)\,,\label{b}\\
& {p}_b\left(T\right)=-2m\gamma L_0\, \frac{\sin\left(\delta_b{b}\left(T\right)\right)}{\delta_b}\,\frac{1}{\gamma^2+\frac{\sin^2\left(\delta_b{b}\left(T\right)\right)}{\delta_b^2}}\, ,
\end{align}
where $b_{o}^{2} = 1 + \gamma^{2}\delta_{b}^{2}$. The black hole horizon lies at $T=0$ where 
$b$ and $p_{b}$ vanish. $T$ decreases as we move to the future from the horizon towards the transition surface $\T$. The area $4\pi p_{c}$  of the horizon in the effective theory is slightly larger than  the classical value $4\pi (2m)^{2}$, but the increase is completely negligible for a macroscopic black hole --{the fractional change is $\sim 10^{-21}$ for a black hole of a million Planck masses and $\sim 10^{-106}$ for a solar mass black hole.} \smallskip

Next, let us consider the exterior region. Now the Killing vector $\partial/\partial x$ is time-like and the metric is given by 
\be \label{extmetric}
\t{g}_{ab} \rmd x^{a} \rmd x^{b} =  - \f{\t{p}_b^2}{\t{p}_c L_o^2} \rmd x^2 + \f{\gamma^{2}\, \t{p}_{c}\, \delta_{\t{b}}^{2}}{\sinh^{2} (\delta_{\t{b}}\t{b})} \rmd T^2+ \t{p}_c\,  \rmd \omega^{2} \, .
\ee
The explicit solutions for $c, p_{c}$ have the same form as (\ref{cpc}), \,with $c, p_{c}$ replaced by $\t{c}, \, \t{p}_{c}$. In solutions for $\t{b}$ and $\t{p}_{b}$, on the other hand, the trigonometric functions of $(b\,\delta_{b})$ are replaced by their hyperbolic analogs:
\ba 
&\cosh \big(\delta_{\t{b} }\,\t{b}(T)\big) = \t{b}_o \tanh\left(\f{1}{2}\Big(\t{b}_o T + 2 \tanh^{-1}\big(\frac{1}{\t{b}_o}\big)\Big)\right), \label{bt}\\
&\t{p}_b(T) = -2m\gamma L_0\,  \f{\sinh \left(\delta_{\t{b}}\, \t{b}(T)\right)}{\delta_{\t{b}}}\, \f{1}{\gamma^{2 }-\f{\sinh^2\left(\delta_{\t{b}}\, \t{b}(T) \right)}{\delta_{\t{b}}^2} }, 
\label{pbt}
\ea
where $\t{b}_{o}^{2} = 1 + \gamma^{2}\delta_{\t b}^{2}$. The horizon lies at $T=0$ where $\t{b}$ and $\t{p}_{b}$ vanish,  and $T$ increases as we go outward to spatial infinity.  Note that the equation of motion (\ref{bt}) implies that $\gamma^{2} - \f{\sinh^2\left(\delta_{\t{b}}\, \t{b}(T) \right)}{\delta_{\t{b}}^2}$ never vanishes in the exterior region where $T\in(0,\infty)$.

\subsection{The near-horizon geometry}
\label{s2.2}

In this sub-section, we will discuss the issue of matching the interior metric with the exterior one. Now, as we approach the horizon (i.e., $T=0$) either from the interior region or the exterior, the metric component  $g_{xx}$ associated with the Killing vector vanishes and $g_{TT}$ diverges (just as $g_{tt}$ vanishes and $g_{rr}$ diverges at the horizon of the Schwarzschild solution). The coordinate $T$ is well-defined; $T=0$ at the horizon, monotonically decreases in the interior and monotonically increases in the exterior. The problem is that the coordinate $x$, the affine parameter of the Killing field, becomes ill-defined  (just as in the Schwarzschild case). The question then is if one can introduce  Eddington-Finkelstein-type  coordinates --bridging the interior and the exterior regions-- in which the metric is well-defined across the horizon. This is indeed possible and  we will now spell out the procedure. As in the Schwarzschild  case, it suffices to focus on the $(x,\, T)$ plane. We can write the exterior and interior metrics, respectively, as 
\be   \rmd \t{S}_{2}^{2} = -\t{f}_{1}(T) \rmd x^{2} + \t{f}_{2}(T) \rmd T^{2}, \quad {\rm and} \quad 
      \rmd S_{2}^{2} = f_{1}(T) \rmd x^{2} - f_{2}(T) \rmd T^{2};
       \ee
where
\be \label{f1f2}  \t{f}_{1}(T) = \f{\t{p}_b^2}{\t{p}_c\, L_o^2}, \,\,\,\t{f}_{2}(T) = \f{\gamma^{2} \t{p}_{c}\, \delta_{\t{b}}^{2}}{\sinh^{2} (\delta_{\t{b}}\t{b})} \quad {\rm and} \quad 
f_{1}(T) = \f{{p}_b^2}{{p}_c L_o^2}, \,\,\,{f}_{2}(T) = \f{\gamma^{2} {p}_{c}\, \delta_{{b}}^{2}}{\sin^{2} (\delta_{{b}}{b})}\, .
\ee
Then, following Eddington and Finkelstein, we are led to define $\t{T}_{\star}$ such that 
\be \rmd \t{T}_{\star} =  \Big({\t{f}_{2}}/{\t{f}_{1}}\Big)^{\f{1}{2}} \rmd T \quad \hbox{\rm and  set} \quad 
v = x + \t{T}_{\star}. \ee
Then, the metric in the exterior region becomes
\be  \rmd \t{S}_{2}^{2} = \t{f}_{1} \rmd v^{2} - 2\, (\t{f}_{1}\, \t{f}_{2})^{\f{1}{2}} \,\,\rmd v \, \rmd T.  \ee
Now, $\t{f}_{1}$ vanishes at $T=0$. Therefore, for the metric to be well-defined at the horizon, we need $\t{f}_{1} \t{f}_{2}$ be smooth and positive in a neighborhood of the horizon. As shown below in Section \ref{s3}, this is indeed the case. In particular, \, $\lim_{T\to 0}  \t{f}_{1} \t{f}_{2} = 4m^{2}$ exactly as in the classical theory; quantum corrections to $\t{f}_{1} \t{f}_{2}$ vanish at the horizon.  Limits of first two derivatives of $\t{f}_{1}$ and $(\t{f}_{1}\, \t{f}_{2})^{\f{1}{2}}$ are also well-defined. For the metric coefficient  $\t{f}_{1}$, they are given by
\be
\frac{1}{1+\epsilon_m}, \quad {\rm and,}\quad \frac{\frac{1}{2}\gamma^{2} \delta^{2}_{\t{b}} + 3 - 4\frac{1-\epsilon_m}{1+\epsilon_m}}{1+\epsilon_m} \, ,
\ee 
respectively, where
\be\label{eq:em}
\epsilon_{m}  = \frac{\gamma^2 L_0^2 \delta_c^2}{64 m^2} = \frac{1}{256}\left(\frac{\gamma \Delta^{\f{1}{2}}}{\sqrt{2\pi}m}\right)^{8/3}\, ,  
\ee
where $\Delta$ is again the LQG area gap. For  the metric coefficient $(\t{f}_{1}\, \t{f}_{2})^{\f{1}{2}}$, they are given by 
 \be 2m \qquad  {\rm and} \qquad  m\big(2+\gamma^{2}\delta_{\t{b}}^{2}\big) \, , \ee
 respectively.  We can repeat the procedure starting from the interior. Again, the limits exist and values of metric coefficients and their first two derivatives match with those coming from the exterior.  Thus, the effective metric is (at least) $C^{2}$ across the horizon $T=0$.  \smallskip

To summarize, although the effective metric was constructed separately in the interior region (using homogeneity of a space-like foliation) and exterior region (using homogeneity of a time-like foliation), it is in fact well defined across the horizon as in the classical case.

\section{The Hawking temperature}
\label{s3}

Let us begin by recalling how Euclidean methods are used to calculate the temperature associated with a Killing horizon (see, e.g., \cite{fulling}). The premise of this calculation is that we have a test quantum field propagating on a given static space-time. Thus, field equations satisfied by the metric do not play any role, nor do considerations of quantum gravity proper. 
\smallskip

Let us first consider a general spherically symmetric, static Lorentzian space-time $(M, \t{g}_{ab})$ with a smooth metric
\be \label{lor-metric}{\t{g}_{ab}} {\rmd }x^{a} \rmd x^{b} = - \t{f}_{1}(r) \, {\rmd} t^{2} + \t{f}_{2} (r) {\rmd} r^{2} + r^{2} {\rmd} \omega^{2},  \ee
where $\rmd \omega^{2} $ again denotes the line element on a unit 2-sphere, and $r \in (r_{\circ}, \infty)$ is the area radius. Suppose the static Killing field $t^{a}$ (with $t^{a}\partial_{a} = \partial/\partial_{t}$) becomes null at $r\!=\!r_{o}$ and is time-like for $r >r_{\circ}$. Then $\t f_{1}(r) \ge 0$ and vanishes only at $r=r_{\circ}$, which is a Killing horizon.%
\footnote{The $(r,\, t)$ chart breaks down at $r=r_{o}$. However, as we saw in Section \ref{s2.2}, conditions that  limits of $f_{1}^{\prime}$ and $f_{1}f_{2}$  do not vanish as $r\to r_{\circ}$ ensure that one can pass to a new chart in which the metric is regular at $r=r_{o}$. In the Euclidean sector discussed below,  the period of the affine parameter $t$ of $t^{a}$ has also to be chosen appropriately. Under these conditions, the breakdown of coordinates $(r,\, t)$  is similar to that of the $(r,\,\theta)$ chart at the origin in flat space. For our purposes, it will suffice to use the $r>r_{\circ}$ part of the manifold and then take the limit as $r\to r_{\circ}$, making sure that the fields in question are such that limits are well-defined.}
Let us now carry out a Wick rotation. Setting $t_{\rm E} = -it$, we obtain a Riemannian metric $\t{\g}_{ab}$:
\be \label{eucl-metric} \t{\g}_{ab} \rmd x^{a} \rmd x^{b} = \t{f}_{1}(r) \, \rmd t_{\rm E}^{2} + \t{f}_{2} (r) \rmd r^{2} + r^{2} \rmd \omega^{2}.  \ee
Since ${\t{\g}_{ab}}$ is Riemannian and $\t{f}_{1}(r_{\circ}) =0$, it follows that the Killing field $t^{a}_{\rm E}$ itself vanishes at $r=r_{o}$. Since a Killing field $t_{\rm E}^{a}$ and its derivative $\nabla_{\!a} t_{\rm E}^{b}$ cannot both vanish simultaneously at any point in space-time \cite{aaam3}, it follows that $\t{f}^{\prime}_{1}(r_{\circ}) \not=0$, where the `prime' denotes derivative with respect to $r$. Furthermore, vanishing of $t_{\rm E}^{a}$ at $r=r_{\circ}$ implies that, under the isometry generated by $t_{\rm E}^{a}$, each point on the 2-sphere  $r = r_{\circ}$ is left invariant and the tangent space at that point is rotated in the $r-t_{\rm E}$ plane. Thus in a neighborhood of $r=r_{o}$, the Killing field  $t_{\rm E}^{a}$ resembles a rotation. 

From now on, it will suffice to focus just on the manifold of orbits of the rotational Killing fields, i.e., the $r-t_{\rm E}$ plane. The `rotational' character of $t_{\rm E}^{a}$ becomes manifest if we set $ R = {(\t f_{1}(r))^{\f{1}{2}}}$ \, so that the metric on the $r-t_{\rm E}$ plane becomes
\be {\t{\g}_{ab}} \rmd x^{a} \rmd x^{b} = R^{2}\, \rmd t_{\rm E}^{2} + 4 \,\f{\t{f}_{1}\t{f}_{2}}{{(\t{f}_{1}^{\prime})^2}} \, \rmd R^{2}\, . \ee
We want to ensure that the metric does not have a conical singularity at $R=0$, i.e, $r=r_{o}$.  Consider then a small circle with radius $\delta R$ and ask that the ratio of its circumference to radius is $2\pi$ in the limit $r\to r_{o}$.  Assuming that $t_{\rm E}$ is periodic with period $\P$ we obtain
\be  \label{nodefect} 2\pi = \lim_{R\to 0}\,\, \f{\rm Circumference}{\rm radius} =  \f{\P\, \t{f}_{1}^{\prime}}{2 (\t{f}_{2} \t{f}_{1})^{\f{1}{2} }}\, . \ee
Note that for the limit of the ratio to exist, the metric has to satisfy the condition $ \lim_{r\to r_{o}}\, \t{f}_{1} \t{f}_{2} > 0$, which \, --unlike the property $\t{f}^{\prime}|_{r_{\circ}} \not=0$\,-- \, is not guaranteed by symmetry conditions.  When this condition is satisfied, our requirement that the metric be regular \, --i.e. have no conical singularity at $r=r_{\circ}$\,-- \, determines 
the period $\P$. One can also express it invariantly using the norm ${\t{\g}}_{ab} \,t_{\rm E}^{a} t_{\rm E}^{b} = \t{f}_{1}$ of the Killing field:
\be \label{period} \P \,=\, \lim_{R\to 0} \, \f{4\pi (\t{f}_{1}\t{f}_{2})^{\f{1}{2}}}{\t{f}_{1}^{\prime}}  \,= \, \lim_{R\to 0}\, \f{4\pi (\t{f}_{1})^{\f{1}{2}} }{||D \t{f}_{1}|| }\, .  \ee
 Thus,  the metric $\t{\g}_{ab}$ is well-defined at  $R=0$, only if  the coordinate $t_{\rm E}$ is periodic with period $\P$. (If $\t{g}_{ab}$ is the Schwarzschild metric, then one recovers the well-known result $\P = 8\pi GM$, where $M$ is the Schwarzschild mass.) Let us consider quantum fields propagating on the Lorentzian space-time  given by the metric (\ref{lor-metric}).  Periodicity in the Euclidean time naturally leads to a thermal state of this field at temperature $T_{\rm H} = \f{\hbar}{K\P}$, associated with the horizon \cite{ggmp}. (Here $K$ is the Boltzmann constant.) \smallskip
 
We can now use this method to calculate the Hawking temperature of the quantum corrected black hole described by the effective metric. The first step is to carry out the Wick rotation of the effective metric in the exterior region.  The effective metric (\ref{extmetric}) has the form:
\be 
\t{g}_{ab} \rmd x^{a} \rmd x^{b} =  - \f{\t{p}_b^2}{\t{p}_c \,L_o^2} \rmd x^2 + \f{\gamma^{2} \t{p}_{c}\, \delta_{\t{b}}^{2}}{\sinh^{2} (\delta_{\t{b}}\t{b})} \rmd T^2+ \t{p}_c  \rmd \omega^{2} \, ,
\ee
where the Killing field is $\partial/\partial x$. Thus, we have to make the following changes in the symbols used in the general discussion above:  $t\to x$ and $r \to T$ and replace the coefficient $r^{2}$ of $\rmd \omega^{2}$ by $\t{p}_{c}$.  The Wick rotated, positive-definite metric in the $(r,\, t)$ plane \, --i.e., now  in the $(T,\, x)$ plane-- \, becomes:
\be \label{2metric}
 {\t{\g}_{ab}} \rmd x^{a} \rmd x^{b}\, =\,   \t{f}_{1} (T) \rmd x^{2} + \t{f}_{2} (T) \rmd T^{2}  \quad {\rm with}
 \quad \t{f}_{1} = \f{\t{p}_b^2}{\t{p}_c\, L_o^2} \,\,\, {\rm and}\,\,\,\,  \t{f}_{2} =  \f{\gamma^{2} \t{p}_{c}\, \delta_{\t{b}}^{2}}{\sinh^{2} (\delta_{\t{b}}\t{b})}\,  .\ee
The horizon is at $T=0$, where $\t{p}_{b}$ and $\t{b}$ vanish in the effective solution.  To check if the 2-metric $\t{\g}_{ab}$ is regular there, we first need to verify that $\t{f}_{1} \,\t{f}_{2} > 0$ in a neighborhood of the horizon.  Eq. (\ref{2metric}) immediately implies 
\be \t{f}_{1} \t{f}_{2} \,=\, \f{\t{p}_{b}^{2} \,\, \gamma^{2}\,\delta_{\t{b}}^{2}}{\sinh^{2} (\delta_{\t{b}} \t{b})\, L_{o}^{2}}. \ee
Thus, we do have $\t{f}_{1}\, \t{f}_{2} > 0$ away from the `horizon' $T=0$. But since $\t{b}$ and $\t{p}_{b}$ both vanish there, we need to evaluate the limit $T\to 0$ to make sure that $\t{f}_{1} \t{f}_{2}$ does not vanish in the limit. It is straightforward to carry out this calculation using the explicit expressions of $\t{p}_{b},\, \t{p}_{c}$ and $\delta_{\t{c}} \t{c}$  in the exterior region given in Section \ref{s2}.  One obtains:
\be \lim_{T\to 0} \, \t{f}_{1}(T) \t{f}_{2}(T)  \,=\,\, {4m^{2}}, \quad \hbox{\rm exactly as in the classical theory.} \ee
(In the classical theory, the product is $1$ in the $(r,\,t)$ coordinates, and since $T$ is related to the Schwarzschild coordinate $r$ via $e^{T} = r/2m$, it is $4m^{2}$ in the $(T,\, x)$ coordinates.)
Thus $f_{1}f_{2}$  is manifestly positive also in the effective theory and we can use (\ref{period}) to calculate the period $\P$.  One obtains:
\be  \P  =  8\pi m (1+\epsilon_{m}), \quad{\rm where} \quad 
\epsilon_{m}  = \frac{1}{256}\left(\frac{\gamma \Delta^{\f{1}{2}}}{\sqrt{2\pi}m}\right)^{8/3}\!\!\!, \quad \hbox{\rm as in Eq. \eqref{eq:em}}.\,   
\ee
This expression is exact for the effective metric under consideration, whence the Hawking temperature of this quantum corrected black hole horizon  is 
\be \label{temp}  T_{\rm H}\,  =\, \f{\hbar}{8\pi K m} \, \f{1}{(1+\epsilon_{m})}, \ee
rather than $\f{\hbar}{8\pi K m}$, the temperature associated with the horizon of the classical Schwarzschild black hole.  The mass dependent term $\epsilon_{m}$ gives the quantum correction to the Hawking temperature. It is very small for macroscopic black holes. For a solar mass black hole it is of the order of {$\sim 4 \times 10^{-106}$}. Indeed, even for a black hole of $\sim 10^{6} M_{\rm Pl}$, the correction is of the order {$10^{-21}$}. (Because there are inherent approximations in arriving at the effective theory, further extrapolation to even smaller black holes would not be appropriate.) 
 
Results in \cite{aos}  showed that the quantum corrections  to various curvature invariants are small near the horizon of macroscopic black holes. The correction $\epsilon_{m}$ to the Hawking temperature provides another facet of that general phenomenon, but one that tests the  near horizon properties of the metric itself. More importantly, nature of this correction brings out the viability of the effective description vis a vis a key quantum property of black holes.\\

\emph{Remark:} There exists in the literature a discussion of the so-called  `dirty' spherical, static black holes \cite{visser}\,  --\!\! `dirty' in the sense that they allow matter fields also outside the horizon. This analysis relates the horizon surface gravity with integrals involving components of the stress-energy tensor of the matter field from the horizon to infinity, and the radius and the energy density at the horizon. Then, assuming that the matter content satisfies the \emph{weak energy condition} and that the space-time metric has the \emph{standard asymptotic fall-off,} one concludes that the surface gravity must satisfy an inequality.  Finally, using the same Euclidean arguments as in the discussion of this section  surface gravity  is related to the Hawking temperature $T_{H}$, to transfer this inequality to the Hawking temperature: 
\be  T_{\rm H} \le \f{\hbar}{\sqrt{4\pi\, A_{H}}} \quad \hbox{\rm with $A_{H}$, the horizon area.} \ee
Does this inequality hold in our case?  Using the fact that the horizon area is given by  $A_{H} = 4\pi \t{p}_{c}\!\mid_{H} \,= 16\pi m^{2} (1+\epsilon_{m})$  and $T_{\rm H}$ is given by (\ref{temp}), it is trivial to verify that it does hold.  It is therefore tempting to inquire if the temperature can be related to the effective matter stress-energy tensor as in \cite{visser}.  However, as was emphasized in  \cite{visser}, that derivation was meant only for classical black holes  since the weak energy condition is violated at the semi-classical level. This is indeed the case of the effective metric now under consideration. Furthermore, as discussed below in Section \ref{s4}, the effective metric is asymptotically flat in a somewhat weaker sense  than required in \cite{visser}.  Therefore the derivation does not go through and the explicit expressions relating surface gravity to matter fields do not hold in our effective geometry.  But since the final result does hold, it may well be that there is a non-trivial generalization of that analysis, applicable to our effective metric $\t{g}_{ab}$.
 
\section{Asymptotic structure }
\label{s4}

As explained in Section \ref{s1}, the  asymptotic behavior of the effective geometry was not analyzed in detail in \cite{aos} largely because (i) it was found that the effective metric is in excellent agreement with the classical theory near horizons of macroscopic black holes, and, (ii)  
{calculations} with MATHEMATICA showed that for curvature scalars the agreement improves as one moves away from the horizons in the outward direction.%
\footnote{For example, even for a black hole with mass as small as $m=  10^{6} \lp$, the square of the scalar curvature $R^{2}$  --which vanishes in the classical theory--  is very small  at the horizon with $R^{2}/K \approx 2.7\times 10^{-11}$\!, where $K$ is the Kretschmann scalar, and the ratio decreases as $m$ increases. $R^{2}$ increases slightly as one moves outward, reaches a maximum at $T\approx 0.13$ and then starts decreasing. (Recall that the horizon is at $T=0$.)}
Therefore it was assumed in \cite{aos} that quantum effects would become totally negligible
as one moves even further away from the horizon. However, it was realized soon thereafter  --and independently pointed out in  \cite{brahmacomment}--  that the situation is not as simple.
Therefore, in this section we will analyze the asymptotic properties of the effective metric in greater detail. We will find that the metric \emph{is} asymptotically flat  in the sense that it approaches a Minkowski metric as $1/r$, but its curvature does not fall off as fast as it does in the standard treatments of asymptotics \cite{geroch-cinn,ah,aa-ein,wald}. This specific asymptotic behavior is of some interest in its own right because, on the one hand it is sufficient to lead to a well-defined Arnowitt-Deser-Misner (ADM)  energy, and on the other hand it brings out the richer structure made available by the stronger fall-off conditions used in the standard treatments. 

\subsection{Approximate expression of the exterior metric }
\label{s4.1}

Recall that the metric in the exterior region is given by (\ref{extmetric}):
\be 
\t{g}_{ab} \rmd x^{a} \rmd x^{b} =  - \f{\t{p}_b^2}{\t{p}_c\, L_o^2} \rmd x^2 + \f{\gamma^{2} \t{p}_{c}\, \delta_{\t{b}}^{2}}{\sinh^{2} (\delta_{\t{b}}\t{b})} \rmd T^2+ \t{p}_c  \rmd \omega^{2}.
\ee
To facilitate comparison with the standard form of the Schwarzschild metric, let us change notation as follows. Set
\be t: = x,\quad  r_{S} := 2m,\quad  r:= r_{S}\,e^T  \quad b_{0} \equiv (1 +\gamma^{2}\delta_{\t{b}}^{2})^{\f{1}{2}}
=: 1 + \epsilon, \ee
and
\be R^{2} := \t{p}_{c} = 4m^2\left( e^{2T}+\frac{\gamma^2L_0^2\delta_{\t{c}}^2}{64m^2}e^{-2T}\right)  \equiv
 r^2\left(1+\frac{\gamma^{2}L_{0}^{2}\delta_{\t{c}}^{2} r_S^2}{16  r^4}  \right)\, . \ee
Note that now the quantum parameters are $\delta_{\t{c}}$ and $\epsilon$. (We switched from $\delta_{\t{b}}$ to $\epsilon$ because it is {$b_{0}=1+\epsilon$}, rather than $\delta_{\t{b}}$, which appears in some of the key equations.) The exterior metric now takes the form:
\be \label{eq:g-tr}\t{g}_{ab} \rmd x^{a} \rmd x^{b} =  \t{g}_{tt}\rmd t^{2} + \t{g}_{rr} \rmd r^{2} +  \t{R}^{2}\,  \rmd \omega^{2} \, . \ee
One can now substitute the solutions (\ref{cpc}), (\ref{bt}) and (\ref{pbt}) in the expression of the metric coefficients and cast them in terms of the newly introduced variables $t, r, r_{S}, \epsilon$. 
The result is:
\begin{equation} \label{gtt}
\t{g}_{tt}= -\left(\frac{r}{r_S}\right)^{2\epsilon}\frac{\left(1-\left(\frac{r_S}{r}\right)^{1+\epsilon}\right)\left(2+\epsilon+\epsilon\left(\frac{r_S}{r}\right)^{1+\epsilon}\right)^{2} \left((2+\epsilon)^{2}-\epsilon^{2}\left(\frac{r_S}{r}\right)^{1+\epsilon}\right)}{16\left(1+\frac{\delta_{\t{c}}^{2} L_0^{2}\gamma^{2}r_S^2}{16  r^4} \right) (1+\epsilon)^{4}}\, ,
\end{equation}
and
\begin{equation} \label{grr}
\t{g}_{rr}= \Big(1+\frac{\delta_{\t{c}}^{2} L_0^{2}\gamma^{2}r_S^2}{16 r^4} \Big)\frac{\Big(\epsilon +\left(\frac{r}{r_S}\right)^{1+\epsilon } (2+\epsilon )\Big)^2}{\Big(\left(\frac{r}{r_S}\right)^{1+\epsilon }-1\Big) \Big(\left(\frac{r}{r_S}\right)^{1+\epsilon } (2+\epsilon )^2-\epsilon ^2\Big)}\, .
\end{equation}
These expressions of the effective metric coefficients are exact; we have not make any approximations or asymptotic expansions. Note that if quantum geometry effects are ignored, i.e. if we set $\delta_{\t{c}} =0$ and $\epsilon=0$, we recover the Schwarzschild metric in the standard form.

We will now simplify these expressions by exploiting the smallness of quantum parameters (for example, $\epsilon \approx  10^{-26}$ for a solar mass black hole) and the property $r/r_{S} >1$ that holds outside the horizon.  Thus, using approximations
\be \Big(2+\epsilon+\epsilon\left(\frac{r_S}{r}\right)^{1+\epsilon}\Big)  \approx  2, \quad \Big((2+\epsilon)^{2}-\epsilon^{2} (\frac{r_S}{r})^{1+\epsilon}\Big) \approx 4, \quad
{\rm and} \quad  \left(1+\frac{\delta_c^{2} L_0^{2}\gamma^{2}r_S^2}{16  r^4} \right) \approx 1,\ee
we obtain:
\be \t{g}_{tt} \approx  -\Big(\frac{r}{r_S}\Big)^{2 \epsilon} \,\Big(1-\left(\frac{r_S}{r}\right)^{1+\epsilon} \Big) \, =: \t{g}^{\circ}_{tt} ,\ee
and analogous approximations yield 
\be \t{g}_{rr} \approx  \Big(1-\left(\frac{r_S}{r}\right)^{1+\epsilon}\Big)^{-1} \,=:\, \t{g}^{\circ}_{rr}
 \qquad {\rm and} \qquad  \t{R}^{2} \approx r^{2}\, .\ee
Thus, in the asymptotic region, the effective metric is extremely well approximated by 
\be \t{g}^{\circ}_{ab} \rmd x^{a} \rmd x^{b} =  \t{g}^{\circ}_{tt}\rmd t^{2} + \t{g}^{\circ}_{rr} \rmd r^{2} +r^{2}\,  \rmd \omega^{2} \, . \ee

Now, these expressions  --particularly the factor $({r}/{r_S})^{2 \epsilon}$ in $\t{g}^{\circ}_{tt}$ that diverges as $r\to \infty$ if $\epsilon\not=0$--  bring out the fact that the effective metric $\t{g}_{ab}$ does not approach the flat metric $\eta_{ab}$  with the line element $ \eta_{ab}\rmd x^{a} \rmd x^{b} =  -\rmd t^{2} + \rmd r^{2} + r^{2} \rmd \omega^{2}$.  This feature led authors of  \cite{brahmacomment} to conclude that the metric is not asymptotically flat and its ADM energy is not well defined.  We will show in the next subsections that these conclusions are unwarranted.

The key point is that while  ${\eta}_{ab}$ is an obvious choice to test asymptotic flatness, it is  not sacrosanct.  Consider, for example, the 2-dimensional metric $\bar{g}_{ab}$ with the line element $\rmd \bar{s}^{2} = - r^{2} \rmd t^{2} + \rmd r^{2}$, which again has $\partial/\partial t$ as the Killing vector, whose norm $\bar g_{tt}$ diverges as $r\to \infty$. Clearly, it does not approach the flat metric $\bar\eta_{ab}$ in the `natural coordinates' with line element $\bar\eta_{ab} \rmd x^{a} \rmd x^{b} = -dt^{2} + dr^{2}$. But not only is  $\bar{g}_{ab}$ asymptotically flat, it is in fact flat! Indeed, $\bar{g}_{ab}$  is just the Minkowski metric in the Rindler wedge.%
\footnote{\label{fn4}  A curved space, 4-dimensional analog of this situation occurs in the case of the Levi-Civita solution to Einstein's equation (known as the `c-metric') \cite{cmetric} that, it turned out, represents the gravitational field of two accelerating black holes \cite{kw,aatd}. In this solution, the norm of the Killing field $\partial/\partial t$ also diverges at spatial infinity, although the metric is asymptotically flat  in the standard sense \cite{aatd}.}
Returning to the effective metric, we will show that although $\t{g}_{ab}$ does not approach the $\eta_{ab}$ defined above, it does approach a Minkowski metric $\eta^{o}_{ab}$  as $1/r$ at spatial infinity.

\subsection{Asymptotic flatness}
\label{s4.2}

The question then is whether the effective metric $\t{g}_{ab}$ is asymptotically flat in spite of appearances. Let us begin by sharpening the notion of {asymptotic} flatness.  We will say a given metric $g_{ab}$ is \emph{asymptotically flat in the elementary sense} at spatial infinity, if there exists a flat metric $\eta_{ab}^{o}$ such that in a Cartesian chart  defined by $\eta^{o}_{ab}$, components of $g_{ab}$ approach the components  of $\eta^{o}_{ab}$ at least as fast as $1/r$,  as $r \to \infty$ keeping $t, \theta,\varphi$ constant  (where $(t, r,\theta,\varphi)$ refer to $\eta^{o}_{ab}$ ).  Thus,  if a given metric $g_{ab}$ is asymptotically flat in this sense, it does approach the Minkowski metric $\eta^{o}_{ab}$. However, as the example of the Rindler wedge explicitly shows, $g_{ab}$ need not approach another Minkowski metric $\eta_{ab}$ on the same manifold. 

Let us return to the asymptotic form $\t{g}_{ab}^{\circ}$ of the effective metric and examine its components in a chart that is better adapted to the issue of asymptotic flatness.  Let us set
\be \label{tau}   \tau =  t\, \Big(\f{r}{r_{S}} \Big)^{\epsilon} \, , \ee 
so that for $\epsilon =0$ --i.e. if we ignore quantum geometry corrections--  $\tau$ reduces to the standard Schwarzschild time coordinate $t$.  Now, as Eq. (\ref{db-dc}) shows, values of  the quantum parameters $\delta_{\tilde b},\, \delta_{\tilde c}$  --and hence of $\epsilon$-- depend on the value of $m$ of the Dirac observable on that solution. Thus, as  one might expect, the transformation $t \to \tau$ depends on the effective metric under consideration. To test asymptotic flatness of a given solution, $\tau$ is defined using the $\epsilon$ of that solution.

It is easy to verify that in the $(\tau, r, \theta,\varphi)$ chart,  the asymptotic metric assumes the form: {
\ba \label{asymmetric}  \t{g}^{\circ}_{ab} \rmd x^{a} \rmd x^{b} &=& \Big( -\rmd \tau^{2} +\rmd r^{2} + r^{2} \rmd \omega^{2} \Big) \,+ \Big(\f{r_{S}}{r}\Big)^{1+\epsilon}\, \rmd \tau^{2} \, +\, 2\epsilon \f{\tau}{r}\, \Big(1 - \big(\f{r_{S}}{r}\big)^{1+\epsilon} \Big) \rmd r \rmd \tau \nonumber \\ 
& -& \Big[ \Big(1 - \f{1}{1-\big(\f{r_{S}}{r}\big)^{1+\epsilon}}\Big) - \epsilon^{2}\, \f{\tau^{2}}{r^{2}}\, \Big(1- \big(\f{r_{S}}{r}\big)^{1+\epsilon}\Big)  \Big]\,  \rmd r^{2}\,  . \ea 
This} form implies that the Cartesian components of $ \t{g}^{\circ}_{ab}$  approach those of 
the flat metric $\t\eta^{\circ}_{ab}$ with line element
\be \label{mink}  \t{\eta}^{\circ}_{ab} \rmd x^{a} \rmd x^{b} = \big( -\rmd \tau^{2} +\rmd r^{2} + r^{2} \rmd \omega^{2} \big) \,  ,\ee
at lease as fast as $1/r$, as $r \to \infty$ keeping $\tau, \theta,\varphi$ constant.  Thus, $ \t{g}^{\circ}_{ab}$  --and hence also the full effective metric $ \t{g}_{ab}$--   is indeed asymptotically flat at spatial infinity in the elementary sense.%
\footnote{Indeed, throughout this calculation we could have worked directly with $\t{g}_{ab}$ in place of $\t{g}^{o}_{ab}$ with minor algebraic changes. But the simple reason behind asymptotic flatness would then have been obscured by the fact that the metric coefficients $\t{g}_{tt}$ and $\t{g}_{rr}$ are much more complicated than $\t{g}_{tt}^{o}$ and $\t{g}_{rr}^{o}$.}
The norm of the Killing field $t^{a}$ of  the effective metric $\t{g}_{ab}$  is of course an intrinsic property of  $\t{g}_{ab}$ --insensitive to our choice of $\eta^{o}_{ab}$-- and it does diverge at spatial infinity. As we mentioned already, this is similar to what happens in the Rindler wedge. But in that case the metric also admits another time-like Killing field with finite norm at spatial infinity. What is the situation in the present case? Since we are now dealing with a curved metric, the result is correspondingly weaker. The time-like vector field $\tau^{a}$ --with unit norm--  (defined by $\tau^{a} \partial_{a} = \partial/\partial \tau$) is now only an \emph{asymptotic} Killing vector of the metric $ \t{g}_{ab}$, exactly as in the case of the Levi-Civita c-metric \cite{cmetric,kw}. \smallskip

However,  asymptotic flatness of $\t{g}_{ab}$ is weaker than in the standard treatments \cite{geroch-cinn,ah,aa-ein,wald} because the usual conditions assume that not only does the  physical metric approach those of a Minkowski metric as $1/r$ but the $n$th derivatives $\mathring\nabla_{a_{1}}  \ldots \mathring\nabla_{a_{n}} \,g_{bc}$  also fall off at least as fast as  $1/r^{n+1}$ for a suitable $n$ ( $\ge 2$). (Here $\mathring{\nabla}$ is the derivative operator of $\eta^{o}_{ab}$.) In our case, while the (Cartesian) components of $\t{g}_{ab}$ do approach those of $\t{\eta}^{o}_{ab}$ as $1/r$ as required,  components of $\mathring{\nabla}_{a} {\t{g}}_{bc}$ do not all fall-off as $1/r^{2}$ because of the presence of $\tau/r$ terms; some fall-off only as $1/r$.  Therefore,  while (Cartesian) components of the curvature tensor do fall off at least as fast as $1/r^{2}$, not all components fall-off as fast as $1/r^{3}$ as in the standard treatments.  Indeed, explicit calculations show that curvature invariants have the following asymptotic behavior:
\begin{equation}
{(\t{R}_{ab}\t g^{ab})^2}=\frac{4 \epsilon ^2 \big((\epsilon
+1)+\big(\frac{r}{r_S}\big)^{-1-\epsilon}\big)^2}{r^4},
\end{equation}
\begin{equation} \label{ricci2}
{\t{R}_{ab} \t{R}^{ab}} =\frac{2 \epsilon ^2 \big(\left(\epsilon ^2+2\right)+2
(\epsilon -1) \big(\frac{r}{r_S}\big)^{-1-\epsilon }+2
\big(\frac{r}{r_S}\big)^{-2-2 \epsilon }\,\big)}{r^4},
\end{equation}
\begin{equation} \label{weyl2}
{\t{C}_{abcd}\t{C}^{abcd}}=\frac{4 \big((\epsilon -2) \epsilon + (\epsilon
-3)\big(\frac{r}{r_S}\big)^{-1-\epsilon }\big)^2}{3 r^4}, \quad {\rm and}
\end{equation}
\begin{equation}
{\t{K}}=\frac{4 \big(\epsilon^2 ((\epsilon -2) \epsilon +3)+ (2 (\epsilon
-1) \epsilon +3) \big(\frac{r}{r_S}\big)^{-2-2 \epsilon }+2
(\epsilon -2) (\epsilon -1) \epsilon
\big(\frac{r}{r_S}\big)^{-1-\epsilon }\big)}{r^4}\, ,
\end{equation}
where $\t{K}$ is the Kretschmann scalar. (Throughout, we have given \emph{only} the leading order terms.)

The question then is if the asymptotic fall-off of $\t{g}_{ab}$ is nonetheless sufficient for the ADM energy to be well-defined. To address this issue, let us first recall a few facts about the ADM energy $E_{\rm ADM}$. In the literature there are several apparently distinct definitions of $E_{\rm ADM}$. The most widely used is the original one involving the Cartesian components of the (asymptotic) spatial metric, and in recent years another, involving the (asymptotic) spatial Ricci tensor, is also often used in the geometric analysis literature \cite{schoen}. While these notions are distinct to start with, it is known that if the metric has the standard fall-off,  all these definitions agree \cite{aaam1,aaam2}.  In the geometric analysis literature it is also known that \emph{some} of these definitions lead to a well-defined ADM energy even if the asymptotic fall-off is significantly weaker than in standard treatments.  Finally, even though the curvature of $\t{g}_{ab}$ falls-off more slowly than in standard treatments, the effective energy density 
{$\rho = \frac{1}{8\pi G} {(\t{R}_{ab} - (1/2) \t{R} \t{g}_{ab})} \h{t}^{a} \h{t}^{b}$} of the space-time has the desired fall-off,
\begin{equation} \label{rho}
\rho = -\frac{\epsilon }{8 \pi G}\,\,\frac{r_S^{1+\epsilon } }{r^{3+\epsilon}},
\end{equation}
making its integral over a spatial slice well-defined. (Here $\h{t}^{a}$ is the unit time-like vector field along $t^{a}$, or equivalently along $\tau^{a}$, since the two are parallel.)  In light of these known results, one might hope that some of the standard definitions may assign a well-defined ADM energy to the effective metric $\t{g}_{ab}$. In Section \ref{s4.3} we will show that this is indeed the case. Furthermore, the calculation is carried out using the notion of time translation provided by the exact Killing field $t^{a}$ of $\t{g}_{ab}$, since the notion of energy refers to the physical time-translation.\\

\emph{Remarks:}  \\
1.  As we emphasized at the end of Section \ref{s4.1}, to test if a given physical space-time $(M, {g}_{ab})$ is asymptotically flat, one has to ask whether the (asymptotic region of the) manifold $M$ \emph{admits} a Minkowski metric $\eta_{ab}^{o}$ to which the ${g}_{ab}$ approaches as $1/r$, and \emph{not} whether ${g}_{ab}$ approaches a \emph{pre-specified} Minkowski metric $\eta_{ab}$ on $M$ as $1/r$. This basic point is overlooked in the note added in the most recent version of \cite{brahmacomment} (version 3). Therefore, we will elaborate on it further.

Consider again the 2-metric $\bar{g}_{ab} \rmd x^{a} \rmd x^{b} = - r^{2} \rmd t^{2} + \rmd r^{2}$ where $r \in [0, \infty)$ and $t\in (-\infty, \infty)$. The norm of the Killing field $\partial/\partial t$ diverges as $r\to \infty$ and so if one were to insist on using the Minkowski metric $\bar\eta_{ab} \rmd x^{a} \rmd x^{b} = -\rmd t^{2} + \rmd r^{2}$ to test asymptotic flatness of  $\bar{g}_{ab}$, one would conclude that it is \emph{not} asymptotically flat at spatial infinity. But let us consider coordinates $\tau, \mathfrak{r}$ defined via
\be \label{rindler}   \tau = r \sinh t, \qquad {\rm and} \qquad  \mathfrak{r} = r\cosh t\, ,   \ee
so,  that we have $\mathfrak{r} \in [0, \infty)$ and $\tau\in (-\infty, \infty)$ with $\mathfrak{r}^{2} - \tau^{2} \ge 0$, explicitly showing that the space-time under consideration is the Rindler wedge. Indeed,  the line element defined by $\bar{g}_{ab}$ becomes $-\rmd \tau^{2} + \rmd \mathfrak{r}^{2}$;  $\bar g_{ab}$ is clearly asymptotically flat at spatial infinity because it is flat! Thus, in place of the naive choice $\bar\eta_{ab}$, we have to use the Minkowski metric $\bar\eta_{ab}^{o} \rmd x^{a} \rmd x^{b} = -\rmd \tau^{2} + \rmd \mathfrak{r}^{2}$ to test asymptotic flatness of $\bar{g}_{ab}$. It is simply incorrect to declare that a metric is not asymptotically flat because it does not approach a \emph{pre-specified} Minkowski metric.%
\footnote{The Levi-Civita c-metric \cite{cmetric,kw} referred to in footnote \ref{fn4} provides a more sophisticated example.  Although the metric is asymptotically flat \cite{aatd}, it took decades to recognize this fact  because the metric is presented in coordinates which obscure the Minkowski metric it approaches.}
This is the simple but key conceptual point that continues to be overlooked in various versions of \cite{brahmacomment}.

Our construction of the Minkowski metric (\ref{mink}) through the introduction (\ref{tau}) of $\tau$ is considered in \cite{brahmacomment} to be ``problematic to examine the asymptotic limit'' because ``in this limit $r\to \infty$,  the new coordinate is finite only when $t \to 0$.''  By this logic, then, the passage (\ref{rindler}) from $t$ to $\tau$ in the Rindler wedge would also have to be regarded as  ``problematic to examine the asymptotic limit'' for exactly the same reason. Hence,  the reasoning of \cite{brahmacomment} would lead to the conclusion that it is incorrect to use  $\bar\eta_{ab}^{o}$ to test asymptotic flatness of the Rindler metric $\bar g_{ab}$ at spatial infinity; we have to stick to $\bar\eta_{ab}$ and conclude that the metric is not asymptotically flat! 

2. In light of Remark 1 it is natural to ask if,  in place of our $\t{\eta}^{\circ}_{ab}$,  we can make an even better choice $\mathring{\eta}_{ab}$ of the Minkowski metric  such that the physical metric $\t{g}_{ab}$ approaches $\mathring\eta_{ab}$ in the standard sense of asymptotic flatness (e.g., as in \cite{geroch-cinn,aa-ein}). The answer is in the negative.%
\footnote{That $\t{g}_{ab}$ is not asymptotically flat in the standard sense was implicit in the first version of the paper. We thank the referee for raising this possibility of existence of another Minkowski metric $\mathring{\eta}_{ab}$, which led to the explicit argument in this remark. Note that some treatments of asymptotic flatness impose a stronger condition requiring that the physical metric should satisfy vacuum equations near infinity (see, e.g. \cite{ah,wald}). This stronger condition is trivially violated by the effective metric, just as it is violated by the Reissner-Nordstrom metric. Therefore in this remark we focus on treatments such as \cite{geroch-cinn,aa-ein} which do not require this stronger condition.} 
For, the standard asymptotic conditions require the Cartesian component of the effective stress-energy tensor $T_{ab}$ to fall-off as $1/\mathring{r}^{4}$, where `Cartesian coordinates' and $\mathring{r}$ refer to $\mathring\eta_{ab}$.  These conditions then imply that the Cartesian component of  the Weyl tensor $C_{abcd}$ would fall-off only as $GM/\mathring{r}^{3}$.  Thus, if $\mathring{\eta}_{ab}$ were to exist, the curvature invariant  $(R_{ab}R^{ab}/ C_{abcd}C^{abcd})$ of the effective metric would have to go to zero at infinity. However, as Eqs. (\ref{ricci2}) and (\ref{weyl2}) show, this is not the case for $\epsilon\not=0$. Thus, the putative $\mathring{\eta}_{ab}$ does not exist; the effective metric $\t{g}_{ab}$ is not asymptotically flat in the standard sense.

\subsection{The ADM energy}
\label{s4.3}

The ADM energy refers to an asymptotic time translation. In the present case, the metric $\t{g}_{ab}$ has an exact time-like Killing field in the asymptotic region and so the energy should refer to this symmetry. Recall, however, that the norm of the Killing field $t^{a}$ --and hence the lapse-- diverges at infinity. Therefore, there is a clear danger that one would obtain an infinite answer using any of the definitions. We will now show that this is not the case.

Let us first consider the standard notion of ADM energy in terms of the spatial metric.  It follows from Eq. {(\ref{eq:g-tr})} that the spatial metric $\t{q}_{ab}$ can be taken to be:
\be \label{q}
 \t{q}_{ab} \rmd x^{a} \rmd x^{b} =  \t{g}_{rr}\rmd r^{2} + \t{R}^{2} \rmd \omega^{2}  \, ,
\ee
with $\t{g}_{rr}$ given by (\ref{grr}). The lapse $N$ adapted to the Killing field $t^{a}$ is given by $N =  \big(-\t{g}_{tt}\big)^{\f{1}{2}}$. The ADM energy is then defined by (see, e.g., \cite{tt}):
\be  E_{\rm ADM}=\lim_{r\to\infty}\frac{1}{16 \pi G} \oint_{r} dS_{d}\, \big(\det \t{q} \big)^{\f{1}{2}} \,\t{q}^{ac} \t{q}^{ bd} \,\big[ N \partial_ {[c} \t{q} _ {b]a} - \left(\t{q}_{a[b} - {\delta}_{a[b}\right) ({\partial}_{c]} N) \big]\, ,  
\ee
where partial derivatives refer to the spatial Cartesian coordinates of $\t\eta^{o}_{ab}$. Substituting for  $\t{q}_{ab}$ from (\ref{q}) we obtain
\begin{equation}
\lim_{{r} \to\infty}\frac{1}{16 \pi G} \oint_{r} dS_{d}\, \big(\det \t{q}\big)^{\f{1}{2}} \,\t{q}^{ac} \t{q}^{bd} \,\big[ N \partial_ {[c} \t{q} _ {b]a}\big]
=\,  \f{m}{G} \, \equiv \, M,
\end{equation}
and 
\begin{equation}
\lim_{{r} \to\infty}\frac{1}{16 \pi G} \oint_{r} dS_{d}\, \big(\det \t{q}\big)^{\f{1}{2}} \,\t{q}^{ac} \t{q}^{bd} \,\big[\left(\t{q}_{a[b} - {\delta}_{a[b}\right) ({\partial}_{c]} N) \big] =0\, ,
\end{equation}
since $\big(\t{q}_{ab} - {\delta}_{ab}\big) \propto D_{a}r  D_{b} r$ and $D_{c} N \propto D_{c} r$ everywhere on the spatial hypersurface. Therefore, we obtain  $E_{\rm ADM}=M$.\medskip

Recall that there is a second expression of the energy in terms of the Ricci tensor $\t{\mathcal{R}}_{ab}$ of the spatial metric $\t{q}_{ab}$ \cite{aaam2}  that is often used in the more recent geometric analysis literature \cite{schoen}:
\begin{equation}
E_{\rm Ricci} =  \lim_{{r} \to\infty}\frac{1}{8 \pi G} \oint_{r} \rmd^{2}V\, \,r\, N \,\, \t{\mathcal R}_{ab} \hat r^a \hat r^b \, ,
\end{equation}
where $\rmd^2 V$ is the area element of the  $r = {\rm const}$  2-sphere of integration, and  $\hat{r}^{a}$ a unit radial vector.  Under standard asymptotic fall-off,  we have $E_{\rm Ricci} = E_{\rm ADM}$ \cite{aaam2}.  What is the situation for the effective metric? An explicit calculation shows that $E_{\rm Ricci}$ is also well-defined. Its value is given by:
\be E_{\rm Ricci} = (1 +\epsilon)\,  M \quad {\rm where}\quad 1+\epsilon = \Big[1+ \Big(\f{\gamma^{2}\Delta}{2\pi m^{2}}\Big)^{\f{1}{3}}\Big]^{\f{1}{2}} \,. 
\ee
As already remarked, for macroscopic black holes $\epsilon$ is extremely small;  for a solar mass black hole, {$\epsilon = 10^{-26}$}. Both these values are essentially the same as the horizon energy, $GE_{H} := \big(\f{A_{H}}{16\pi}\big)^{\f{1}{2}}$, defined by the geometrical radius of the horizon, but differ by a quantum correction: 
\be E_{H} = E_{\rm ADM} \,  \Big(1+\frac{\gamma^2L_0^2\delta_c^2}{64m^2}\Big)^{\f{1}{2}}  \equiv   E_{\rm ADM}\,\big( 1 + \epsilon_{m}\big)\, . \ee
As we saw in Section \ref{s2.1}, for a solar mass black hole this quantum correction is {$\sim 10^{-106}$!} Thus, for macroscopic black holes three quite different notions of energy associated by the effective space-time turn out to have essentially the same value. Note that the fact that the answers are not \emph{exactly} the same is not surprising, but it is instructive.  It is not surprising because notions that all agree in a given theory often differ when we pass to a more general theory with richer physics. We encounter numerous examples of this phenomena in planetary astronomy as we move from Newtonian gravity to general relativity, or in atomic physics as we transit from non-relativistic quantum mechanics to relativistic quantum mechanics and then to quantum field theory.  The phenomenon is instructive because: (i)  it brings out the fine balance struck by the standard notion of asymptotic flatness, where these apparently disjoint expressions agree \emph{exactly} \cite{aaam2}, and,  (ii) it invites us to better understand the physics behind the differences in quantum corrections, which would in turn  strengthen our intuition on the quantum nature of geometry.

To summarize, even though the $\t{g}_{tt}$ component of the asymptotic form of the effective metric $\t{g}_{ab}$ in the $(t,r,\theta,\varphi)$ chart diverges as $r\to \infty$, we showed explicitly that $\t{g}_{ab}$ is in fact asymptotically flat in the precise sense defined in Section \ref{s4.2}; it is just that the Minkowski metric it approaches is not the `obvious' Minkowski metric associated with the $(t,r,\theta,\varphi)$ chart. In addition, although the asymptotic fall-off of $\t{g}_{tt}$ is weaker than that in standard treatments in that several components of the space-time curvature have weaker fall-off than in the standard context, the ADM energy associated with the time translation symmetry $t^{a}$  is nonetheless well-defined. These results could be taken as an indication that, even though the norm of the time-like Killing field grows unboundedly as we approach spatial infinity, the effective metric $\t{g}_{ab}$ manages to capture correct physics. \smallskip

We will conclude this discussion with a few remarks.\\
1. The surprising element in  the asymptotic properties of $\t{g}_{ab}$ is that the norm of its time translation Killing field $t^{a}$ grows as $\big(r/r_{S}\big)^{\epsilon}$ even though $\t{g}_{ab}$ approaches a flat metric $\t{\eta}^{o}_{ab}$ as $1/r$.  This unforeseen behavior of $t^{a}$ can be `understood' as follows. Consider two situations: (i)  a static black hole with horizon radius $r_{H}$, and, (ii) a black hole surrounded by a thick shell of matter between $r_{1}$ and $r_{2}$ with $r_{H} < r_{1} < r_{2}$. Then, if the shell has positive energy density the norm of the Killing field at some given radius $r = r_{o} >r_{2}$ in the second space-time would be smaller than that for the first space-time.  Reciprocally, if the energy density of the shell were negative, the norm in the second space-time at $r= r_{o} >r_{2}$ would be greater than in the first space-time without any matter shell. In the effective space-time under consideration, as we saw in Eq. (\ref{rho}), we have an effective energy density $\rho$ in the asymptotic region, which falls off as $\rho \propto \epsilon\, /r^{3+\epsilon}$ but is \emph{negative}.  Had we switched off quantum gravity effects, we would have $\epsilon=0$ and $\rho =0$. Thus, it  is because the quantum corrected space-time has a negative effective energy density that the norm of the static Killing field at any given radius $r_{o}$ in the asymptotic region is greater in the quantum corrected space-time than in its classical counterpart.

2. The ADM energy $E_{\rm ADM}$ we calculated refers to the time translation $t^{a}$ of the physical metric. Had we calculated the generator of the asymptotic time translation $\tau^{a}$, we would have obtained zero. This is because a time evolution that corresponds to the change in $\tau$ by a finite amount $\tau_{o}$ amounts to a change in the affine parameter of $t^{a}$ by $r^{-\epsilon} \tau_{o}$, which vanishes at infinity. Thus, this would be a `bubble time evolution' vis a vis the affine parameter $t$ of $t^{a}$, whence we would expect its generator to vanish, just as it does.

3. Even though the effective metric passes these tests, it may still be unsettling that if one goes further and further in the asymptotic region, the metric components in the $(t,r,\theta,\varphi)$-chart deviate more and more from those of the Schwarzschild metric of mass $M$. Note, however, that since we are comparing two \emph{different} metrics, there is some ambiguity in saying what one means by `the same point' in these different space-times. Instead of identifying points with the same values of $(t,r,\theta, \varphi)$ in the effective space-time ($\epsilon \not =0)$ and the Schwarzschild space-time ($\epsilon =0$), if we first introduce the coordinates $(\tau, r, \theta, \varphi)$ in each of these space-times, and then identify points with same values of \emph{these} coordinates, the two metrics \emph{would} approach each other (see Eq. (\ref{asymmetric})). 

But one could still insist on using $(t,r,\theta,\varphi)$ in this comparison, say, by appealing to the fact that these coordinates have a natural geometrical interpretation. Suppose we do so and ask when the deviations would become important. Then, for the deviation to be, say, 10\%, for a solar mass black hole one would have to move away from the horizon $\sim10^{10^{24}} r_{S}$ because {$\epsilon \approx 10^{-26}$} in this case. Note that this distance is \emph{overwhelmingly} larger than the radius of observable universe which is  $\sim 5 {\rm Gpc} \approx 10^{22} r_{S}$ (since $r_{S} \approx 3 {\rm km}$ in this case). Even for a black hole of $10^{6}{\rm M_{\rm Pl}}$, one would have to move away $\sim 10^{1158} \lp$, while $5 {\rm Gpc} \approx 10^{61}\lp$. We can also ask a reciprocal question. Suppose for a solar mass black hole we move away from the horizon by $5 {\rm Gpc}$, i.e. all the way to the edge of the observable universe. How big a deviation would there be for $g_{tt}$ from the Schwarzschild metric? It would be of the order  $10^{-23}$.  Thus, even if we insist on using the $(t,r,\theta,\varphi)$-chart  in the comparison, from a physical view point, there is little reason to be concerned about viability of the effective metric. 

\section{Discussion}
\label{s5}

The issue of singularity resolution has drawn considerable attention in all major approaches to quantum gravity. As we explained in Section \ref{s1},  in contradistinction to views sometimes expressed in the literature on string theory, and on the `weak gravity conjecture', the viewpoint in LQG is that it is the \emph{quantum nature of geometry} that  leads to a natural resolution of space-time singularities. This viewpoint is realized in a concrete manner in LQC, where the issue has been discussed extensively from various angles (see, e.g., \cite{ps,iaps}). In particular, the effective equations have been derived quite systematically starting from the dynamics of sharply peaked \emph{quantum} states in LQC \cite{jw,vt,asrev}. For black holes, on the other hand, the situation is less satisfactory. While effective equations have been introduced taking inspiration from LQC, so far there is no systematic derivation starting from (a symmetry-reduced version of) LQG.  Rather, just as effective equations provided a powerful tool in an earlier phase of LQC,  providing  guidance for full quantum theory beyond the FLRW models, the hope is that effective equations would serve the same purpose for the ongoing efforts (see, e.g.\cite{gop,gp-hawking,cgop,gp-crit,hhcr,bcdhr}) to construct a more complete quantum description of the singularity resolution for black holes.  

The approach to effective dynamics presented in \cite{aoslett,aos} was in this spirit. Therefore, the focus was largely on overcoming the limitations of the previous effective descriptions of the Schwarzschild interior (see, e.g.   \cite{ab,lm,bv,dc,ck,cgp,bkd,djs,cs,oss,cctr,yks}) by using a new prescription --with plausible theoretical underpinnings-- to fix the quantum parameters $\delta_{b},\, \delta_{c}$.  This strategy provided a satisfactory effective description of the quantum extension of the space-time beyond the classical singularity in the sense that it succeeded in overcoming all known limitations of the previous approaches to the Schwarzschild interior. However, being an effective description, its scope only covers macroscopic black holes, say with $M >10^{6} M_{\rm Pl}$; 
it  does not incorporate the  full quantum gravity effects that are needed to handle Planck mass ones. Nonetheless, its attractive features  --e.g., on absolute bounds on curvature-- suggest 
that it probably captures some of the essential features of the mechanisms that are expected to lead to the resolution of black hole singularities in full LQG. Because of these features, it has already served as a point of departure for some generalizations --e.g., to include physical matter fields in the model \cite{zz}, and to extend it to dynamical situations involving collapse \cite{rama}.

The approach in \cite{aoslett,aos} also provided a proposal to extend previous investigations of the symmetry reduced Schwarzschild interior to the exterior region by exploiting  the fact that the exterior geometry of the Schwarzschild solution also admits 3-surfaces with homogeneous geometries, although they are now time-like. However, in \cite{aoslett,aos} the effective geometry of the exterior region was not investigated in as much detail as that of the interior. Therefore, in this note we analyzed some features of the exterior. First of all,  since the effective metric was constructed separately in the interior and exterior region, a priori it is not obvious that it would match smoothly across the horizons that separate these regions.  We provided Eddington-Finkelstein type coordinates in Section \ref{s2.2} to explicitly show that this is the case. More importantly, having a candidate effective metric for the exterior region, one could perform a Wick transform and compute the quantum gravity correction to the Hawking temperature. We calculated this correction exactly in Section \ref{s3}.  It is negligibly small for macroscopic black holes, as hoped from the general LQG perspective. 

In Section \ref{s4} we investigated the asymptotic properties of the effective geometry. We showed that the metric does approach a Minkowski metric as $1/r$ in a precise sense and the ADM energy is well-defined and agrees --within small quantum corrections-- with the horizon energy, computed using the horizon area. However, we also found that the curvature of the effective metric goes to zero at a rate that is slower than in the standard treatments of asymptotic flatness. As we discussed in Section \ref{s4.3},  for the macroscopic black holes considered in this paper, this feature has no observational consequences within our cosmological radius.   Nonetheless, it could well be that there is a more astute choice of the quantum parameters $\delta_{b}, \delta_{c}$ --or a modification of the treatment of the exterior effective geometry-- that yields the standard fall-off of curvature at infinity while retaining the merits of the current  effective descriptions. While interesting ideas in this direction are available (see, e.g., \cite{modesto2}), their conceptual underpinning is not transparent. Perhaps the detailed analysis of the asymptotic structure presented in this note will clarify the type of modifications that are likely to lead to the desired results both in the interior and in the exterior, and serve as guidelines for new proposals that are also conceptually well-motivated.

A limitation of the effective equations discussed in this note is that because they are obtained starting with a symmetry reduced theory, it is not known if there is a 4-dimensional covariant action whose symmetry reduction yields these equations.  However, this is a limitation only of the present status of the program: contrary to the suggestion made in \cite{mb-comment}, there is no viable obstruction.  Because of spherical symmetry, one may be tempted to use known results in  2-dimensional metric theories of gravity to probe this issue. But note that there is no a priori requirement that the full theory, without symmetry reduction, must be a metric theory. Indeed,  already in the simpler homogeneous isotropic cosmology (Friedmann-Lemaitre-Robertson-Walker models), it was initially far from being obvious that the (now widely used) effective equations of LQC arise from the symmetry reduction of a covariant 4-dimensional field theory. It turned out that there \emph{is}  such a covariant action and, furthermore, the full (symmetry-unreduced) theory has the same degrees of freedom as general relativity. However,  it is based on a Palatini action, that uses both a metric $g_{ab},$ {\it and} an independent affine connection $\nabla_{a}$ (that is distinct from the metric connection $\nabla^{(g)}_{a}$) \cite{os}.%
\footnote{Furthermore, in \emph{general} Palatini theories, action can contain traces of products of  the Ricci tensor $R_{a}{}^{c} = R_{ac}g^{cb}$ and still the theory has no additional degrees of freedom if $R_{[ab]} =0$. These are not known to be equivalent to scalar tensor theories \cite{olmo}. Therefore arguments that appeal to scalar tensor theories \cite{mb-comment} are also inconclusive.} 
The situation may be similar with the effective equations used in this note. Another possibility is that  the desired (symmetry-unreduced) covariant equations do exist but they involve variables that are non-local in the metric. For example, in semi-classical gravity describing 1+1 dimensional black holes  including back reaction, one often solves for the conformal factor $e^{2\rho}$ that relates the physical metric to the flat metric (so $2\, \Box \rho = - R$, the physical scalar curvature), and derivatives of $\rho = - \f{1}{2}\Box^{-1} R $ feature in the quantum corrected equations  which are fully covariant (see, e.g., \cite{strominger-review}). Finally, the symmetry-unreduced, general covariant equations could involve additional \emph{non-dynamical} variables --e.g., gauge fields which are non-dynamical in 1+1 dimension. Indeed,  a decade long puzzle \cite{DVV,GV} that equations for a 1+1 stringy black hole were not of the form of a dilaton gravity theory was resolved by introducing such fields \cite{grumiller}.  It would be of considerable interest if one could use an avenue along these lines to construct a covariant action for the current effective equations (or, for a more satisfactory modification thereof). Such a discovery would enable one to address several important questions --e.g., the issue of stability, details of gravitational collapse,   Hawking radiation and its back reaction, in the regime in which the black hole remains macroscopic, and space-time dynamics beyond the Page time. As we mentioned above, there are several works in progress within LQG to better understand the  physics of black holes beyond effective descriptions \cite{gop,gp-hawking,cgop,gp-crit,hhcr,bcdhr,han}.  Just as the present effective equations can provide guidance to these investigations, results obtained in the broader context of LQC could also suggest strategies to embed effective equations in a more complete framework.

\section*{Acknowledgments}  We would like to thank Tommaso De Lorenzo and especially Amos Ori and Parampreet Singh for stimulating discussions, Jim Bardeen for useful correspondence, and an anonymous referee for informing us of Ref. \cite{visser}. This work was supported in part by the  NSF grants PHY-1505411 and PHY-1806356,\, grant UN2017-9945 from the Urania Stott Fund of the Pittsburgh Foundation, the Eberly research funds of Penn State, and by Project. No. FIS2017-86497-C2-2-P of MICINN from Spain. J.O. acknowledges the Operative Program FEDER 2014-2020 and the {\it Consejer\'ia de Econom\'ia y Conocimiento de la Junta de Andaluc\'ia}.

\end{document}